\documentclass[aps,showpacs,twocolumn,superscriptaddress]{revtex4}

\usepackage{graphicx}
\usepackage{amsmath}
\usepackage{amsfonts}
\usepackage{amssymb}
\usepackage{bm}
\usepackage{psfrag}

\begin{document}

\title{Time-controlled charge injection in a quantum Hall fluid}

\author{T. Jonckheere}
\affiliation{ Centre de Physique Th\'eorique, Case 907 Luminy, 13288 Marseille cedex 9, France}
\author{M. Creux}
\affiliation{ Centre de Physique Th\'eorique, Case 907 Luminy, 13288 Marseille cedex 9, France}
\affiliation{Universit\'e de la M\'edit\'erann\'ee, 13288 Marseille cedex 9, France}
\author{T. Martin}
\affiliation{ Centre de Physique Th\'eorique, Case 907 Luminy, 13288 Marseille cedex 9, France}
\affiliation{Universit\'e de la M\'edit\'erann\'ee, 13288 Marseille cedex 9, France}

\date{\today}

\begin{abstract}
We consider the injection of a controlled charge from a normal metal into an edge state of the fractional quantum Hall effect,
with a time-dependent voltage $V(t)$. Using perturbative calculations in the tunneling limit, and
a chiral Luttinger liquid model for the edge state, we show that the electronic correlations prevent the charge
fluctuations to be divergent for a generic voltage pulse $V(t)$. This is in strong contrast with the case
of charge injection in a normal metal, where this divergence is present. We show that explicit formul{\ae} for the mean
injected charge and its fluctuations can be obtained using an adiabatic approximation, and that non perturbative results
can be obtained for injection in an edge state of the FQHE with filling factor $\nu=1/3$.
Generalization to other correlated systems which can be described with the Luttinger liquid model,
 like metallic Carbon nanotube, is given.
\end{abstract}


\maketitle

\section{Introduction}
It is now well known that the fluctuations of electric current contain valuable information 
both on the discreetness of the charge and on the quantum properties of 
transport~\cite{Blanter_Buttiker00,Martin04,Reznikov_Heiblum95,Kumar_Saminadayar96}.
Many studies of these fluctuations, both experimental and theroretical,
 have been done on systems in a stationary regime, with constant or time-periodic voltage biases. 
We consider here a non-stationnary problem, where a voltage pulse is used to inject a given charge in a conductor,
and call this process ``time-controlled charge injection''. Such a real-time transfer of charge might prove
of great interest for applications, for example as a tool for the transfert of information.
The current fluctuations also play an important role in the problem of time-controlled charge injection, as 
these fluctuations should be made as low as possible to transfer the charge as precisely as possible.

An interesting area to study time-controlled charge injection consists of conductors with strongly correlated
electrons. Indeed, in these systems, the elementary excitations are collective electronic excitations, and may have
a charge $e^*$ which is only a fraction of the ``elementary'' charge $e$. We will be particularly interested in
edge states of the fractional quantum Hall effect (FQHE) for Lauglhin fractions,
 where elementary excitations  have a charge $e^* = e/(2 n+1)$, with $n$ an integer 
(the most accessible charge being $e^* = e/3$)~\cite{Wen92,Saminadayar_Glattli97,Picciotto_Reznikov97}.
 This problem is interesting on its own from a theoretical perspective, 
as it allows to see how a system with strongly correlated electrons behaves
in a non-stationary setup. It is also interesting for potential experimental applications, as the injection of 
a well controlled charge, e.g. a {\it unique} electron, in an edge state of the FQHE is an important experimental challenge,
which could prove an useful tool in the quantum information domain for example.

In the case where electrons are uncorrelated (normal metal conductors), this problem 
has been studied by Levitov and coworkers~\cite{Levitov96}.
Defining the Faraday flux $\phi = e/\hbar \, \int_{-\infty}^{\infty} \, dt V(t)$, they have shown that the mean 
transmitted charge $\langle Q \rangle$ is simply proportional to the flux (Ohm's law),
 $\langle Q \rangle \sim \phi$, but that the charge fluctuations $\langle Q^2 \rangle$ are in
general logarithmically divergent, except for ``integer'' values of the flux $\phi = 2 \pi n$ 
(with $n$ an integer) where these fluctuations are finite. They have related this behavior 
to the Anderson orthogonality catastrophe~\cite{Anderson67}.The goal of this report is to study this problem in the case
 where the charge is transfered into an edge state of the FQHE, or more generally in a chiral Luttinger liquid.

The setup of this paper is as follows. The system is described in section~\ref{sec:description}. In section~\ref{sec:pert}, 
we consider the case of the pertrubative regime, where the tunneling between the normal metal and the edge state is low. 
The convergence of the mean charge and of its fluctuations is studied, and explicit formul{\ae} are given within the adiabatic
approximation, whose validity is confirmed by numerical calculations. In section~\ref{sec:allorders}, we give non-perturbative
results for the special case of electron tunneling in an edge state of the FQHE with filling factor $\nu=1/3$.
 Finally, in section~\ref{sec:others}, some perspectives are discussed, like the relevance of our results for other systems, 
and conclusions are given.

\section{Description of the system}
\label{sec:description}
The system we consider is composed of a normal metal (non-interacting electrons, with Fermi liquid behavior) 
close to an edge state of a 2D electron gas in the FQHE regime. A voltage pulse $V(t)$
is applied between the two conductors, which leads to the tunneling of electrons. The FQHE regime
is characterized by the filling factor $\nu = 1/m$, where $m$ is an odd integer. Note that the
case $m=1$ correspond to the integer quantum Hall effect, where there are no correlations between
electrons and the edge state describe a Fermi liquid.
 We expect thus to recover, when we take $\nu=1$, the results of Levitov et al.  
\begin{figure}
\centerline{ \includegraphics[width=9.cm]{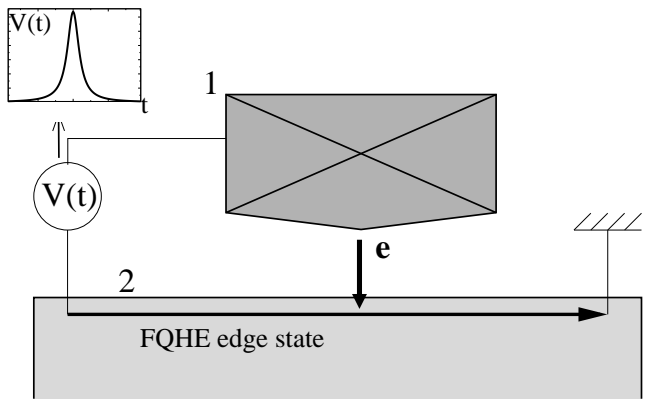}}
\caption{The setup: electron tunneling between a usual conductor 
and an edge state of a 2D electron gas in the fractional quantum Hall effect (FQHE) regime, induced
by a time dependent voltage $V(t)$.}
\label{fig:setup}
\end{figure}

We call $\psi_1$ ($\psi_2$) the electron annihilation operator at the tunneling point in 
conductor 1 (conductor 2, the edge state)- see Fig.\ref{fig:setup}. Using Peierls' substitution to have the 
voltage pulse as a vector potential only, we have the following tunneling Hamiltonian between the two conductors:
\begin{equation}
\label{eq:Hdef}
H_T(t) = \hbar v_F \, \Gamma e^{i \phi(t)} \psi_1^{\dagger}(t) \psi_2(t) + \mbox{h.c.}
\end{equation}
where $\Gamma$ is the tunneling amplitude and $\phi(t) = e/\hbar \int_{-\infty}^{t} dt' V(t')$ is
the time-dependent Faraday flux. Similarly, the tunneling current is given by:
\begin{equation}
\label{eq:Idef}
I_T(t) = i \, e \, v_F \, \left(\Gamma  
 e^{i \phi(t)} \psi_1^{\dagger}(t) \psi_2(t) - \mbox{h.c.} \right)
\end{equation}
The quantities we need to calculate are the mean transmitted charge $\langle Q \rangle$:
\begin{equation}
\label{eq:Q}
\langle Q \rangle = \int_{-\infty}^{\infty} \!\!\!\!\! d\tau \left \langle I_T(\tau) \right \rangle
\end{equation}
and its fluctuations $\langle \Delta Q^2 \rangle = \langle Q^2 - \langle Q \rangle ^2 \rangle$:
\begin{multline}
\label{eq:Q2}
\left \langle \Delta Q^2 \right \rangle =
  \left \langle \int_{-\infty}^{+\infty} \!\!\!\!\! dt  \; I(t) \; \int_{-\infty}^{+\infty} \!\!\!\!\! dt' \;
                   I(t') \right \rangle  - \left \langle Q \right \rangle^2\\
  =  \int_{-\infty}^{+\infty} \!\!\!\!\! dt \, d\tau \; \Big( \left \langle I(t) I(t+\tau) \right \rangle 
    - \left \langle I(t) \right \rangle \left \langle I(t+\tau)\right\rangle \Big)\\
  =  \int_{-\infty}^{+\infty} \!\!\!\!\! dt \, d\tau S(t,t+\tau)
\end{multline}  
The system being out of equilibrium,
we use the Keldysh formalism, introducing a time contour going first from $-\infty$ to $+\infty$ 
(upper branch, $\eta = +1$) and then going back from $+\infty$ to $-\infty$ (lower branch, $\eta=-1$),
and using the time ordering operator $T_K$ along this contour~\cite{Chamon_Freed95}.

\section{Perturbative results}
\label{sec:pert}

\subsection{Formalism}

In this section, we will calculate the mean transmitted charge and its fluctuations in the tunneling regime, where $\Gamma \to 0$.
This allows us to get the results by calculating only the lowest order in the tunneling amplitude $\Gamma$. Note that $\Gamma \to 0$
ensures the system is in the tunneling regime for any finite value of $V(t)$. Indeed, for electron tunneling from a normal 
metal to an edge state of the FQHE with filling factor $\nu=1/(2 p +1)$, one has in the tunneling limit the 
tunneling current $I \sim \Gamma^2 |\omega_0|^{2/\nu -1}$, where
$\omega_0$ is proportionnal to the applied voltage $V$. As $2/\nu -1 > 0$, it is clear that when
$V(t) \to 0$ the tunneling current goes to 0 and can thus be calculatd perturbatively. Note that the situation 
would be different for tunneling of fractionaly charged excitations between two edge states of the same FQHE fluid,
 where $V \to 0$ brings the system out of the tunneling regime

 As we are considering the system in the tunneling regime,
 we can restrict ourselves to the lowest order contribution in the tunneling amplitude $\Gamma$. 
At order $\Gamma^2$, we have for the mean transmitted charge:
\begin{multline}
\left \langle Q \right \rangle  =  \frac{-i}{2 \hbar} \int_{-\infty}^{+\infty} \!\!\!\!\! d\tau \!\! \sum_{\eta \eta_1 = \pm 1} \! \eta_1 
        \!\!\int_{-\infty}^{+\infty} \!\!\!\!\! dt_1 \left \langle T_K I_T(\tau^{\eta}) H_T(t_1^{\eta_1}) \right \rangle
\end{multline}
Using Eqs.~(\ref{eq:Hdef}) and (\ref{eq:Idef}), with standard properties of the Keldysh Green functions,
 and parity properties, we can write this expression as:
\begin{multline}
\label{eq:Icalc}
\left \langle Q \right \rangle  = 
 -2 e \, v_F^2 \Gamma^2  \int_{-\infty}^{+\infty} \!\!\!\!\! d t \int_{-\infty}^{+\infty} \!\!\!\!\! d\tau \;
 \mbox{Im} \left(G_1(t) G_2(t) \right)
\\  \sin\left(\phi(\tau + t/2) - \phi(\tau - t/2) \right)
\end{multline}
where $G_i(t)$ is the standard Green function for conductor $i$ ($i=1,2$): $G_i(t) = \left \langle
T \psi_i^{\dagger}(0) \psi_i(t) \right \rangle$.
Similarly, we get for the charge fluctuations (Eq.~(\ref{eq:Q2})):
\begin{multline}
\label{eq:Scalc}
\left \langle \Delta Q^2 \right \rangle  =
2 e^2 v_F^2 \,\Gamma^2 \int_{-\infty}^{\infty} \!\!\!\!\! dt \int_{-\infty}^{\infty} \!\!\!\!\! d\tau \;
 \mbox{Re} \left(G_1(t) G_2(t) \right) \\
\left[ \cos\left(\phi(\tau + t/2) - \phi(\tau - t/2) \right) - 1 \right]
\end{multline}
Note that the term ``-1'' next to the cosine has been introduced to regularize the expression. This 
regularizing term is needed because we have permuted the order of  the $t$ and $\tau$ integrals, 
and is not needed if the $t$ integral is performed before the $\tau$ integral, as $\int dt \mbox{Re}(G_1(t) G_2(t)) =0$.

\subsection{Convergence of the integrals}
Eqs.~(\ref{eq:Icalc}) and (\ref{eq:Scalc}) will allow us to study the properties of the charge injection.
On these formul\ae, we see that both the injected charge and its fluctuations are obtained with two elements:
 the Green functions product $G_1(t)G_2(t)$, which contains all the information about the two conductors,
 and a kernel obtained by integrating over $\tau$ a function of the Faraday flux $\phi(t)$, which contains all the
information about the voltage pulse $V(t)$.  To study the convergence/divergence of the time integrals for the
charge and its fluctuations, we need the time behavior of these two elements. 
Note that as we consider tunneling through a single point contact, the normal metal  can be mapped~\cite{Chamon_Fradkin97} to a chiral
Luttinger liquid with parameter $\nu =1$. The Green functions at zero temperature are simply~\cite{Chamon_Freed95}:
\begin{equation}
\label{eq:G1G2}
G_1(t) = \frac{1}{2 \pi a}(1 + i v_F t /a)^{-1}  \quad \quad  G_2(t) = \frac{1}{2 \pi a}(1 + i v_F t/a)^{-1/\nu}
\end{equation}
where $a$ is a short length cutoff, and $\nu=1/m$ the filling factor of the FQHE conductor.
Introducing $K=(1/2)(1 + 1/\nu)$, with $K$ an integer, we see that the large time behavior of
the real and imaginary part of $G_1(t) G_2(t)$ is:
\begin{equation}
\mbox{Im}\left(G_1(t) G_2(t)\right)  \sim t^{-(2 K + 1)} \quad \quad 
\mbox{Re}\left(G_1(t) G_2(t)\right)  \sim t^{-2 K}
\end{equation}
(the imaginary part does not contain a $t^{-2 K}$ term  as this term includes a $\sin(\pi K)$ factor which is zero).
We now turn to the large time behavior of the kernels involving the flux  $\phi(t)$. For the mean charge,
it is given by $B_1(t) = \int_{-\infty}^{\infty} d\tau \sin\left(\phi(\tau + t/2) - \phi(\tau-t/2)\right)$.
By hypothesis, the voltage pulse $V(t)$ is important only in a finite time domain, of width $\Delta t$.
This means that for $t \gg \Delta t$, and  for a $\tau$ interval of the order of $t$, we have:
\begin{multline}
\phi(\tau + t/2) - \phi(\tau-t/2) = \frac{e}{\hbar} \int_{\tau-t/2}^{\tau+t/2} \!\!\!\! V(t) \\ \simeq 
\frac{e}{\hbar} \int_{-\infty}^{\infty} \!\!\!\! V(t) = \phi  .
\end{multline}
We have thus
$B_1(t) \simeq \sin(\phi) t + C_1$ for $t \to \infty$, where $C_1$ is a constant.
Similarly, we have for the kernel of the charge fluctuations, 
$B_2(t) =  \int_{-\infty}^{\infty} d\tau [ \cos\left(\phi(\tau + t/2) - \phi(\tau-t/2)\right) -1]$,
$B_2(t) \simeq (\cos(\phi)-1) t + C_2$. Both the kernels $B_1(t)$ and $B_2(t)$ have thus a linear dependence
in $t$ for large $t$, except for the special values of the flux $\phi = 2 \pi n$ (with $n \in \mathbb{N}$) where
they are constant for large $t$.

Combining the large time behavior of the Green functions and of the kernels, we see that the integrand
for the mean charge (Eq.~(\ref{eq:Icalc})) behaves for large $t$ as:
\begin{equation}
\label{eq:Iconv}
B_1(t) \, \mbox{Im}\left(G_1(t)G_2(t)\right) \simeq \sin(\phi) t^{-2 K} + C_1 \, t^{-2 K +1}
\end{equation}
while for the charge fluctuations (Eq.~(\ref{eq:Scalc})) we have:
\begin{equation}
\label{eq:Sconv}
B_2(t) \, \mbox{Re}\left(G_1(t)G_2(t)\right) \simeq (\cos(\phi)-1) t^{-2 K+1} + C_2 t^{-2 K}
\end{equation}
Let us first check that this is compatible with the known results for non-interacting electrons.
In this case, one has $\nu=1$ and thus $K=1$.
 We see then that the mean charge integral is always converging, while the charge fluctuations integral
has a logarithmic divergence, except for $\phi = 2 \pi n$, and we recover thus the results of Levitov 
et al.~\cite{Levitov96}. Turning now to interacting electrons, one has $\nu = 1/m$ with $m>1$ an odd integer,
and thus $K$ is an integer with $K>1$. In this case, we see that the mean charge integral is as before always converging,
and that the charge fluctuations integral is also always converging, independently of the value of the flux $\phi$ !
This means that, because of the electronic correlations, the divergence of the charge fluctuations is 
removed.
 
\subsection{Explicit formul{\ae} and the adiabatic approximation}
It is possible to go further and to get explicit formul{\ae}  for the integrals of the mean injected charge and its fluctuations.
For simplicity, we will restrict ourselves to the case $\nu = 1/3$, but the results shown here can be extended to any value
of $\nu = 1/(2 n+1)$. For $\nu = 1/3$, the Green function product is:
\begin{eqnarray}
\label{eq:Greenproduct}
\mbox{Re}\left(G_1(t)\,G_2(t)\right)  & = & 
     \frac{1}{4 \pi^2 a^2} \frac{1 - 6 (v_F t/a)^2 + (v_F t/a)^4}{(1+(v_F t/a)^2)^4}  \nonumber \\
\mbox{Im}\left(G_1(t)\,G_2(t)\right)  & = & 
 \frac{1}{4 \pi^2 a^2} \frac{4 (v_F t/a) (1 - (v_F t/a)^2)}{(1+(v_F t/a)^2)^4}
\end{eqnarray}
Let us first consider the integral for the mean transmitted charge (Eq.~(\ref{eq:Icalc})), which involves the imaginary
part of the Green function product. On Eq.~(\ref{eq:Greenproduct}), we see that this part is 
important in a time domain of the order of $a/v_F$, and then quickly decreases to 0 for $t \gg a/v_F$ as $t^{-5}$. 
For the $t$ integral giving the mean transmitted charge, we can thus consider that times up to $t \sim a/v_F$ only contribute
importantly to the integral. In the kernel $B_1(t)$, time $t$  appears 
in the bounds $\tau \pm t/2$ of the $V(t')$ integral in the sinus.
As the short time cutoff $a/v_F$ is much smaller that the typical time of variation of $V(t')$, the function $V(t')$ can be considered as 
constant in this integral, giving $\sin(e/\hbar \, V(\tau) \, t)$. The mean charge is then:
\begin{multline}
\label{eq:Iadia1}
\left \langle Q \right \rangle =
 (-1) 2 e \, v_F^2 \Gamma^2 
\int_{-\infty}^{+\infty}\!\!\!\!\! d\tau 
\int_{-\infty}^{+\infty}\!\!\!\!\! dt \;  \sin\left(\frac{e}{\hbar} \,  V(\tau) \, t \right)
\\ \mbox{Im}\left(G_1(t) G_2(t)\right)
\end{multline}
The $t$ integral in~(\ref{eq:Iadia1}) is simply the formula for the mean tunneling current 
when a constant voltage $V(\tau)$ (with a given $\tau$) is applied.
We have thus performed an {\it adiabatic} approximation, valid because of the rapid decrease of the Green function product. 
Performing the time integral, using Eq.~(\ref{eq:Greenproduct}), we get the result:
 \begin{equation}
\label{eq:Iadiares}
\left \langle Q \right \rangle = 
\frac{e \Gamma^2 a^2}{2 \pi \, 3! v_F^2} \int_{-\infty}^{+\infty}\!\!\!\!\! d\tau \;\left(\frac{e}{\hbar} \, V(\tau)\right)^3 
\end{equation}
We see that the mean transmitted charge is not proportional to the total flux $\phi$, but rather to the integral of the cube
of the voltage pulse $V(t)$. This means that, for a voltage pulse of given shape whose flux is varied by an overall scale factor only,
the mean transmitted charge varies as the cube of the flux $\phi$.
This behavior is a consequence of the non-linear relation between voltage and current for tunneling in a Luttinger liquid.
The presence of the cutoff $a$ in the formula is typical of electron tunneling in a Luttinger liquid.
Note that for another filling factor $\nu = 1/m$, the result would be proportional to the integral
of $(V(\tau))^m$.  

Turning now to the charge fluctuations, one could expect to obtain similar results: from Eq.~(\ref{eq:Greenproduct}), we see that
the real part of the product of the Green functions is important for $t \sim a/v_F$ only, and decreases rapidly to 0 
as $t^{-4}$. However, when $V(\tau)$ is not large enough the adiabatic approximation breaks down, because 
$\int_{-\infty}^{+\infty} \! dt  \, \mbox{Re}(G_1(t)G_2(t)) = 0$, which means that the contribution of the time interval
up to  $t \sim a/v_F$ may vanish.
 We have thus to distinguish between two regimes, small flux $\phi \ll 1$ and large flux $\phi$.

For the small $\phi$ regime let us consider a voltage pulse $V(t) = \phi V_1(t)$,
 where $V_1(t)$ is of fixed shape and unit flux ($\int dt \, V_1(t) = 1$),
 and  $\phi \ll 1$. The cosine in the kernel of the charge
fluctuations $B_2(t)$ can then be developped at second order, giving
 $\frac{(-1)}{2} \phi^2 \left( \int_{\tau - t/2}^{\tau + t/2} dt' V_1(t') \right)^2$.
When we vary the flux $\phi$ corresponding to the voltage pulse $V(t)$, we see that the kernel $B_2(t)$, and thus the 
charge fluctuations, varies as $\phi^2$. The charge fluctuations are thus proportional
to  $\phi^2$ for small $\phi$, contrarily to the mean transmitted charge which varies as $\phi^3$.

In the other regime, with a larger flux $\phi$, we can use the same adiabatic approximation as for the mean charge. 
We get:
\begin{multline}
\label{eq:Sadia1}
\left \langle \Delta  Q^2 \right \rangle = 2 e^2 v_F^2\Gamma^2
\int_{-\infty}^{+\infty}\!\!\!\!\! dt \;  \mbox{Re} \left(G_1(t) G_2(t) \right) \\
 \left( \int_{-\infty}^{+\infty}\!\!\!\!\! d\tau  \left[ 
       \cos\left(\frac{e}{\hbar} V(\tau) \, t \right) - 1 \right] \right) 
\end{multline}  
Permuting the $t$ and $\tau$ integrals (the ``-1'' term does not contribute then), we see that for $V(t)$ large enough the 
variations of the cosine  are rapid enough to get a non-zero integral on the domain where the Green functions are important ($t \sim a$).
Performing the $t$ integral, we get for the charge fluctuations, for large $\phi$:
\begin{equation}
\label{eq:Sadiares}
\left \langle \Delta Q^2 \right \rangle =
 \frac{e^2 \Gamma^2 a^2}{2 \pi \, 3! v_F^2} \int_{-\infty}^{+\infty}\!\!\!\!\! d\tau \left( \frac{e}{\hbar} V(t) \right)^3
\end{equation}
In this regime, we recover for the charge fluctuations a behavior similar to the one of the mean charge, with a dependence as $\phi^3$ 
for a voltage pulse of given shape. Comparing Eqs.~(\ref{eq:Iadiares}) and (\ref{eq:Sadiares}), we see that the charge 
fluctuations are simply $e$ times the mean charge. This means that for larger flux, although the system is non-stationary, the charge
fluctuations have a poissonian character, as is common in time-independent problems.

\begin{figure}
\centerline{\includegraphics[width=8.cm]{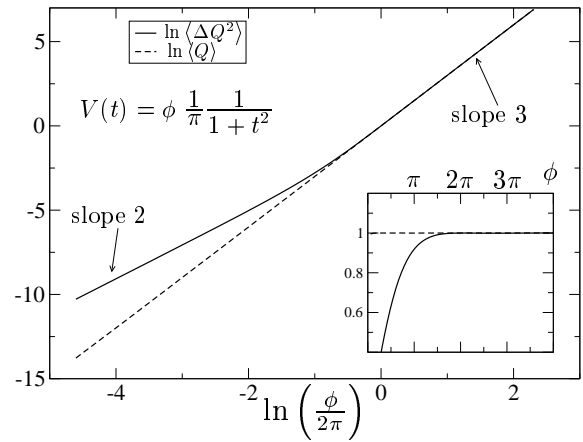}}
\caption{Results of the numerical integration of Eqs.~(\ref{eq:Icalc}) and (\ref{eq:Scalc}), for a pulse
of Lorentzian shape ($V(t) = \phi \;  (1/\pi) (1+t^2)^{-1}$). Mean transmitted charge (dashed curve) 
and its fluctuations (full curve) as a function of $\phi$, on a log-log plot. The mean charge has a slope 3, while
the charge fluctuations has a slope 2 for small $\phi$ and a slope 3 for large $\phi$. Inset: ratio between the results of
the adiabatic approximation (Eqs.~(\ref{eq:Iadiares}) and (\ref{eq:Sadiares})) and numerical integration of
 Eqs.~(\ref{eq:Icalc}) and (\ref{eq:Scalc}), for the mean charge (dashed line) and its fluctuations (full line). The
adiabatic approximation is clearly valid for the mean charge, while it is valid for the charge fluctations when $\phi \gtrsim 2 \pi$. }
\label{fig:numres}
\end{figure}

To confirm the results obtained with the adiabatic approximation, we have performed a numerical integration
 of Eqs~(\ref{eq:Icalc}) and (\ref{eq:Scalc}). The results for the case of a Lorentzian voltage pulse, 
$V(t) =\phi \; (1/\pi) (1+t^2)^{-1}$ are shown on Fig.~\ref{fig:numres}.
 On this figure, it is clear that the mean transmitted charge behaves as $\phi^3$ for all $\phi$, while the
charge fluctuations behaves as $\phi^3$ for large $\phi$, but as $\phi^2$ for small $\phi$. 
The inset of Fig.~\ref{fig:numres} shows the ratio between the numerical integrations and the results of the 
adiabatic approximation, Eqs.~(\ref{eq:Iadiares}) and (\ref{eq:Sadiares}); we see that the adiabatic approximation
gives excellent results for $\langle Q \rangle$ for all fluxes $\phi$, while it gives excellent results for 
$\langle \Delta Q^2 \rangle$ when $\phi \gtrsim 2 \pi$. Numerical integration with other shapes of voltage pulses (not shown)
gives similar results.

\section{Non perturbative results}
\label{sec:allorders}
In the previous section, we have shown that, except for the charge fluctuations at small $\phi$,  
we obtain a very good approximation of the exact results by 
using an adiabatic approximation, where the transmitted charge due to the voltage pulse $V(t)$ is computed
by integrating over $t$ the stationnary current $I_{st}$ due to $V=V(t)$.
As the adiabatic approximation is related to the rapid decrease of the Green function product $G_1(t) G_2(t)$,
and as higher orders of the tunneling current imply higher powers of this product, we expect the adiabatic 
approximation to be valid in the non-perturbative regime. We will thus compute non-pertubative results for
the mean charge and its fluctuations starting from non-pertrubative results for the stationnary 
tunneling current and noise. Because this method implies the calculation of stationnary
current only, it is much simpler than the full calculations.   

Non-perturbative results for the stationnary current are known in the case of tunneling
of electrons from a normal metal to an edge state of the FQHE with filling factor $\nu=1/3$. Indeed,
for tunneling properties, this system is equivalent to the tunneling of electrons between two edges states
of the same quantum Hall fluid with $\nu =1/2$, as shown in~\cite{Sandler_Chamon99}. This equivalence can
be guessed from Eq.~(\ref{eq:G1G2}): the Green function product is $G_1(t) G_2(t) = (2 \pi a)^{-2} (1+i v_Ft/a)^{-(1+1/v)}$,
which for $\nu = 1/3$ is the same as $(2 \pi a)^{-2}(1+i v_F t/a)^{-1/v'} (1+iv_F t/a)^{-1/v'}$ with $v'=1/2$.
In~\cite{Sandler_Chamon99}, non-perturbative results for the tunneling current and noise are obtained for this system,
as it is linked by the duality symmetry to the tunneling of fractionnaly charged excitations (with $e^* = e/2$)
between two $\nu=1/2$ edge states, which can be treated non-perturbatively by refermionization.

Using results of~\cite{Sandler_Chamon99}, we have for the stationnary tunneling current, $I_{st}$, corresponding to 
voltage $V$:
\begin{equation}
I_{st} =  \frac{e^2}{4 \pi \hbar} V - \frac{ e \,v_F}{4 \pi   \, \Gamma a} \mbox{Arctan}\left( \frac{e V}{\hbar v_F}  \, \Gamma a \right)
\end{equation}
This expression shows there are two extreme regimes for the current. For $V$ or $\Gamma \to \infty$, the tunneling is made through
a barrier of large transparency, and the current goes to the maximal value $e^2 V/(2 h)$. On the opposite, for $V$ or 
$\Gamma \to 0$, the barrier transparency goes to 0, and we recover the perturbative results with $I \sim \Gamma^2 V^3$. To get
the mean charge transmitted with a voltage pulse pulse $V(t)$, we now simply integrate over $t$ the current $I_{st}$ 
corresponding to $V(t)$. This gives:
\begin{equation}
\langle Q \rangle =  \frac{e}{4 \pi} \phi - \int_{-\infty}^{\infty}\!\!\!\!\! dt \;  \frac{e\, v_F}{4 \pi \,\Gamma a} \mbox{Arctan}
\left( \frac{e V(t)}{\hbar v_F}  \Gamma a  \right)
\end{equation}  
In the tunneling regime, $\Gamma \to 0$, we recover the expression~(\ref{eq:Iadiares}) obtained perturbatively, where
the mean charge varies as $\phi^3$. Note that we
obtain the same expression for the limit of very small voltage $V(t)$ for any given $\Gamma$ if $\Gamma V(t) \ll 1$ for all $t$.
 On the opposite, in the limit
of a large $\Gamma$ or large $V(t)$, the charge becomes simply proportionnal to the flux $\phi$. The systems shows thus
a crossover between $\langle Q \rangle \sim \phi^3$ at low $\phi$ and $\langle Q \rangle \sim \phi$ at large  $\phi$,
the position of the crossover being a monotonic decreasing function of $\Gamma$.

We now turn to the charge fluctuations. Using again the results of~\cite{Sandler_Chamon99}, we get the following
expression for the non-perturbative stationnary current noise $S_{st}$ due to voltage $V$, with $x= (e V/ \hbar v_F)  \,\Gamma a$:
\begin{equation}
S_{st} = \frac{e^2 v_F}{4 \pi  \,\Gamma a} \left[ \mbox{Arctan}(x) -
                           \frac{ x}{1+x^2} \right]
\end{equation}
For $V$ or $\Gamma \to 0$, we recover the tunneling regime, with $S=e \, I \sim \Gamma^2 V^3$. As for the current, 
we obtain the charge fluctuations $\langle \Delta Q^2\ \rangle$ due to a voltage pulse $V(t)$ by integrating over $t$ the   
noise $S_{st}$ corresponding to $V=V(t)$. Defining $x(t) = (e V(t) /\hbar v_F) \, \Gamma a$, we  get
\begin{equation}
\left \langle \Delta Q^2 \right \rangle = \frac{e^2 \, v_F}{4 \pi  \,\Gamma a} \int_{-\infty}^{\infty}\!\!\!\!\! dt \;
  \left[ \mbox{Arctan}(x(t)) -
                           \frac{ x(t)}{1+x(t)^2} \right]
\end{equation}
There is also a crossover in the behavior of $\langle \Delta Q^2 \rangle$ as a function of $\phi$: when $\phi$ is small , we
recover the results of Eq.~(\ref{eq:Sadiares}) with $\langle Q^2 \rangle \sim \phi^3$.
 As we know from the previous section, the adiabatic approximation
does not reproduce in this case the correct $\phi^2$ behavior for $\phi \to 0$.
 When $\phi$ is large it is more difficult to get analytically the
behavior of $\langle \Delta Q^2 \rangle(\phi)$; numerical integration for different types of $V(t)$ shows that 
$\langle \Delta Q^2 \rangle \sim \phi^{1/n}$ when $V(t) \sim t^{-n}$ for $t \to \infty$. This dependance on the asymptotic properties
of $V(t)$ comes from the fact that for large $\phi$, noise comes mainly from the $t$ regions where
$\Gamma V(t)$ is small (otherwise  the barrier has a high transparency and noise is small), which are simply
the tails of the voltage pulse.

\section{Perspectives and conclusions}
\label{sec:others}
The results we have obtained so far are valid for electron injection in any chiral Luttinger liquid, and can be
applied to other systems than edge states of the fractional quantum Hall effect. In particular, 
the conduction electrons in a single-wall metallic Carbon nanotube can be described in terms of
different modes of chiral Luttinger liquids. The electronic Green function in such a nanotube 
can be shown to be~\cite{Egger_Gogolin98,Crepieux_Guyon03}:
\begin{equation}
G_2(t) = (2 \pi a)^{-1} (1+i v_F t/a)^{-\xi} \mbox{  with } \xi = \frac{3}{4} + \frac{1}{8} \left(g + g^{-1}\right)
\end{equation}
where $g$ is the parameter characterizing the interactions in the nanotube (typical experimental values
are in the range $[0.2,0.3]$). Note that $\xi >1$, but $\xi$ is not an integer. We can then repeat, 
{\it mutatis mutandis}, the same reasoning as in section~\ref{sec:pert} to study the mean charge and its fluctuations.
We define  $K=(1/2)(1 + \xi)$, with $K>1$. As $K$ is not an integer, the large time behaviors of the real and imaginary part
of $G_1(t)G_2(t)$ are here the same:
\begin{equation}
\mbox{Im}\left(G_1(t) G_2(t)\right)  \sim t^{-2 K} \quad \quad 
\mbox{Re}\left(G_1(t) G_2(t)\right)  \sim t^{-2 K}
\end{equation}
Repeating the reasoning that leads to Eqs.~(\ref{eq:Iconv}) and (\ref{eq:Sconv}), we
see that because $K>1$, both the mean charge and its fluctuations are finite, independently
of the value of the flux $\phi$ (and thus of $V(t)$). As for an edge state of the FQHE, the electrons
correlations in a nanotube prevent the divergence of the charge fluctuations for non-stationary injection.

Our results can also be of interest for the very general case of non-stationnary electron tunneling between two normal metals,
when an Ohmic impedance is present in the system. Indeed, it is known that a mapping exists between a coherent one-channel conductor coupled
to an Ohmic environment (the dynamical Coulomb blockade problem~\cite{Ingold_Nazarov92})
 and a Luttinger liquid with an impurity~\cite{Safi_Saleur04}. Our results suggest then that the divergence of
charge fluctuations as predicted by Levitov and coworkers~\cite{Levitov96} are suppressed by the coupling to an Ohmic environment,
as this coupling can be mapped to electron interaction leading to Luttinger liquid behavior. 

To conclude, we have shown that for the non-stationnary charge injection from a normal metal into a chiral Luttinger liquid,
using a voltage pulse $V(t)$, the electronic correlations in the Luttinger liquid
prevent the charge fluctuations from being divergent for a generic voltage pulse. 
In the perturbative regime with respect to  the tunneling amplitude, we have shown that explicit formul{\ae} for the mean injected charge
 and its fluctuations can be obtained using an adiabatic approximation. We have identified when this adiabatic approximation breaks down,
and shown that when it is valid, it leads to a Schottky-like relation bewteen the charge and its fluctuations.
 Finally,  we have  obtained non-perturbative results for the special case of charge injection in an edge state of the
 FQHE with filling factor $\nu = 1/3$. 
 Centre de Physique Th\'eorique is UMR 6207 du CNRS, associated to Universit\'e de la M\'edit\'erann\'ee, Universit\'e 
de Provence and Universit\'e de Toulon.

\end{document}